\documentclass[reprint,amsmath,amssymb,aps]{revtex4-1}
\usepackage{graphicx}
\usepackage{dcolumn}
\usepackage{bm}
\usepackage[utf8]{inputenc}
\usepackage{float}
\usepackage{subfigure}
\usepackage{xcolor}
\usepackage[export]{adjustbox}[2011/08/13]

\begin{document}
	
	\title{Topological characterization of world cities}
	\author{G. S. Domingues$^1$}
    \author{F. N. Silva$^1$}
    \author{C. H. Comin$^2$}
    \author{L. da F. Costa$^1$}
    
    \affiliation{$^1$S\~ao Carlos Institute of Physics, University of S\~ao Paulo, S\~ao Carlos, SP, Brazil}
    \affiliation{$^2$Department of Computer Science, Federal University of S\~ao Carlos, S\~ao Carlos, SP, Brazil}

	\date{\today}
	
	\begin{abstract}
The topological organization of several world cities are studied according to respective representations by complex networks. As a first step, the city maps are processed by a recently developed methodology that allows the most significant urban region of each city to be identified.  Then, we estimate many topological measures on the obtained networks, and apply multivariate statistics and data analysis methods to study and compare the topologies.  Remarkably, the obtained results show that cities from specific continents, especially Anglo-Saxon America, tend to have particular topological properties. Such developments should contribute to better understanding how cities are organized and related to different geographical locations worldwide. 
	\end{abstract}
	
	\maketitle

	\section{Introduction}

Recently, network science emerged as a new scientific field incorporating methods to characterize and model a wide range of systems represented by networks. Differently from traditional graph theory, which usually deals with simpler or more regular graphs; the topologies studied in network science often originate from complex systems, which, in turn, give rise to particularly intricate graphs potentially comprising a large number of elements and relationships. Such a kind of more ellaborate graph is usually called a complex network.

Researchers from several areas have successfully used complex networks to model the most varied types of complex effects~\cite{ALBERT}, such as the relationship between knowledge areas; communication and telephony networks;  electric power transmission systems; citations in scientific papers; financial market; interaction between proteins; or even links between web pages~\cite{COSTA2011}.

The mathematical modeling of urban areas is an important tool for the analysis of the structure and dynamics of cities~\cite{STRANO2013}. Many aspects of cities have been investigated in the literature, including the analysis of the distribution of buildings and streets~\cite{jacobs2016death,jackson1987crabgrass}, population density~\cite{forrester1970urban}, traffic dynamics~\cite{herman1979two}, etc. Among such studies are those focused on the analysis of the structure of cities comprised of streets and crossings, which can therefore be represented in terms of networks~\cite{porta2006network}.

Previous studies on characterizing the structure of cities by using complex networks have provided important contributions to the understanding and potential improvement of aspects such as traffic system~\cite{echenique2005dynamics}, city growth~\cite{barthelemy2009co}, city planing~\cite{costa2010efficiency,crucitti2006centrality} and others~\cite{barthelemy2016structure}. In these networks, street crossings and endings are represented by nodes, while the streets themselves constitute the edges. Through these approach, it is possible to derive an overview of the topological structure of a city and then study its respective properties.
 	
A major problem in applying network science to study urban topology is that it is particularly difficult to determine the respective boundaries. Although these are, in principle, related to administrative regions, which are typically available in the databases, a more precise definition has to be able to consider   proximity of other urban nucleous, and presence of geographical features such as rivers, mountains, etc~\cite{COMIN2016}.
	
A method has been developed to identify and isolate the real urban region of interest of cities~\cite{COMIN2016}. This method is based on the analysis of the spatial density of vertices in the respective networks.  Associated with pre-established parameters, this analysis allows to isolate a region of interest, separating it from adjacent streets with lesser importance that could otherwise camouflage the limits of the urban region. Topological properties only make sense when the relevant urban region is considered, since otherwise we would have several vertices, loops and paths biasing the characteristics of the graph representing the city. 

In this paper, we adopt the previously developed framework for urban city identification~\cite{COMIN2016}, allowing us to perform a large-scale search in order to obtain a more representative databased of world cities.  After acquiring topological measures of the cities in the database, we characterize their distribution in the attribute spaces using multivariate methods such as PCA (Principal Component Analysis)~\cite{Jolliffe2002}.  In particular, we aim at identifying groups of cities from several continents, as well as their particular characteristics, leading to better understanding of the organizational structure of cities.  The obtained results indicate specific properties of cities from different continents, in particular from Anglo-Saxon America.
	
This work begins by presenting the methodology used in determining the set of cities and in complex network modeling, as well as the methodology previously developed to delimit the urban region of a city. Next, the adopted attributes and measures are applied to several cities and the results are discussed.

\section{Methodology}
	
\subsection{City database}
\begin{figure}[!htbp]
		\includegraphics[width=0.9\columnwidth]{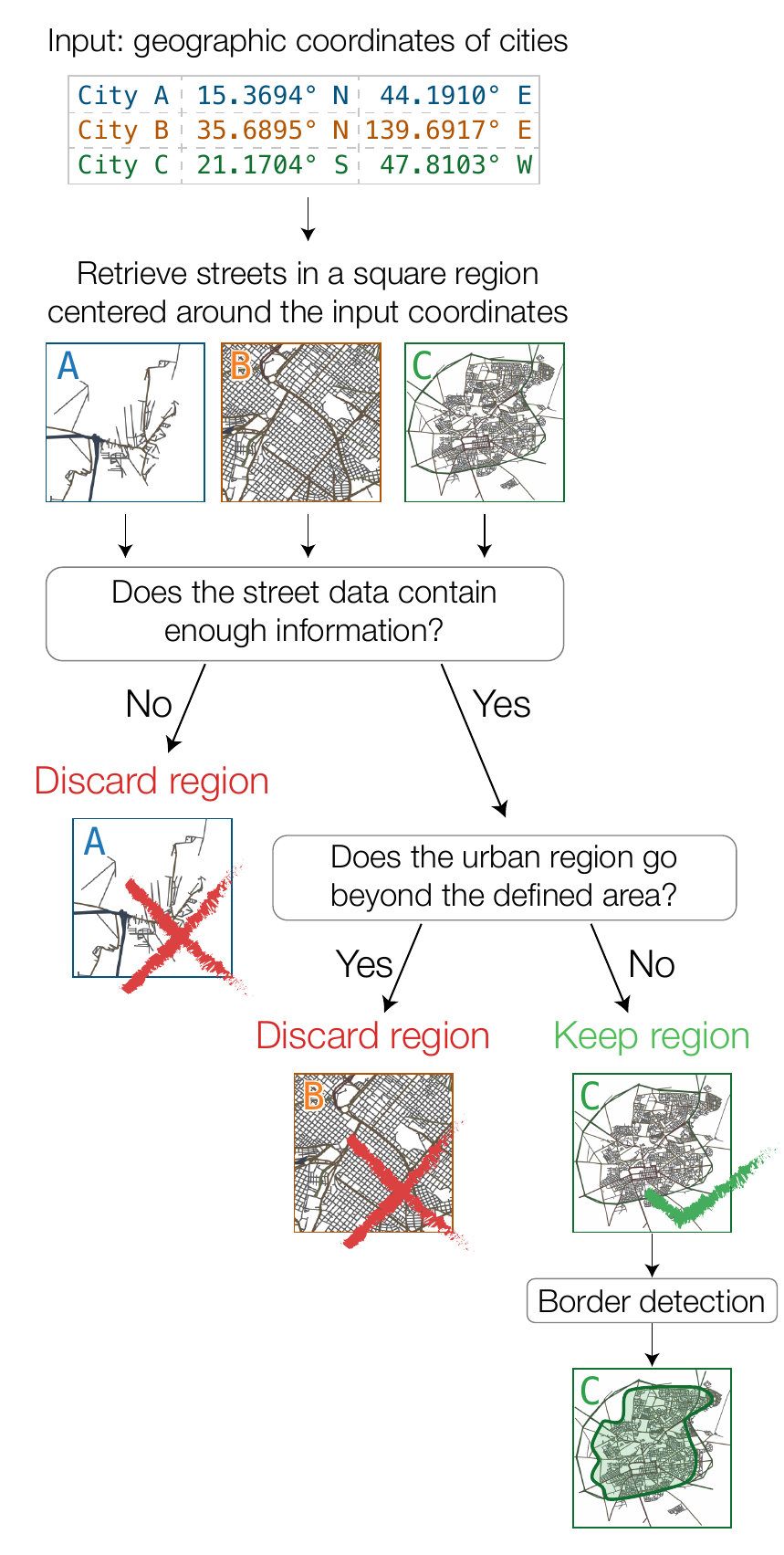}
		\caption{Framework for filtering cities for further analysis.}
		\label{f:filter_methodology}
	\end{figure}

Since it is problematic to compare the topology of cities with largely different sizes, in this work we only consider cities having between 100000 and 600000 inhabitants. The information about city population was collected using the Mathematica function \emph{CityData}. The street network of cities having populations within the aforementioned interval were downloaded from the OpenStreetMaps dataset~\cite{OSM2016}. A sample of this set was analyzed qualitatively in order to develop an exclusion criteria, which is needed by the urban extraction technique~\cite{COMIN2016}.  The database filtering procedure is illustrated in Figure~\ref{f:filter_methodology}.  The subsequent detection of the borders of the selected cities is also included for the sake of completeness.  The filtering starts by receiving the coordinates of the given cities (A, B, C, $\ldots$) and, for each of these coordinates, all streets comprised in the square window of size $L \times L$ are retrieved from the OpenStreetMaps dataset.  If the number of streets is deemed to be enough for the analysis, the region is further checked to ensure most of the streets do not have direct continuations outside the considered window.  Both these checkings are performed interactively by an operator.  The regions that pass all these verifications are then considered for subsequent analysis, which involves the application of the method for the identification of the areas of interest (described below).  As a result, a set of 1150 cities from different countries was selected.

\subsection{Network modeling}
\begin{figure}[!htbp]
		\includegraphics[width=1.0\linewidth]{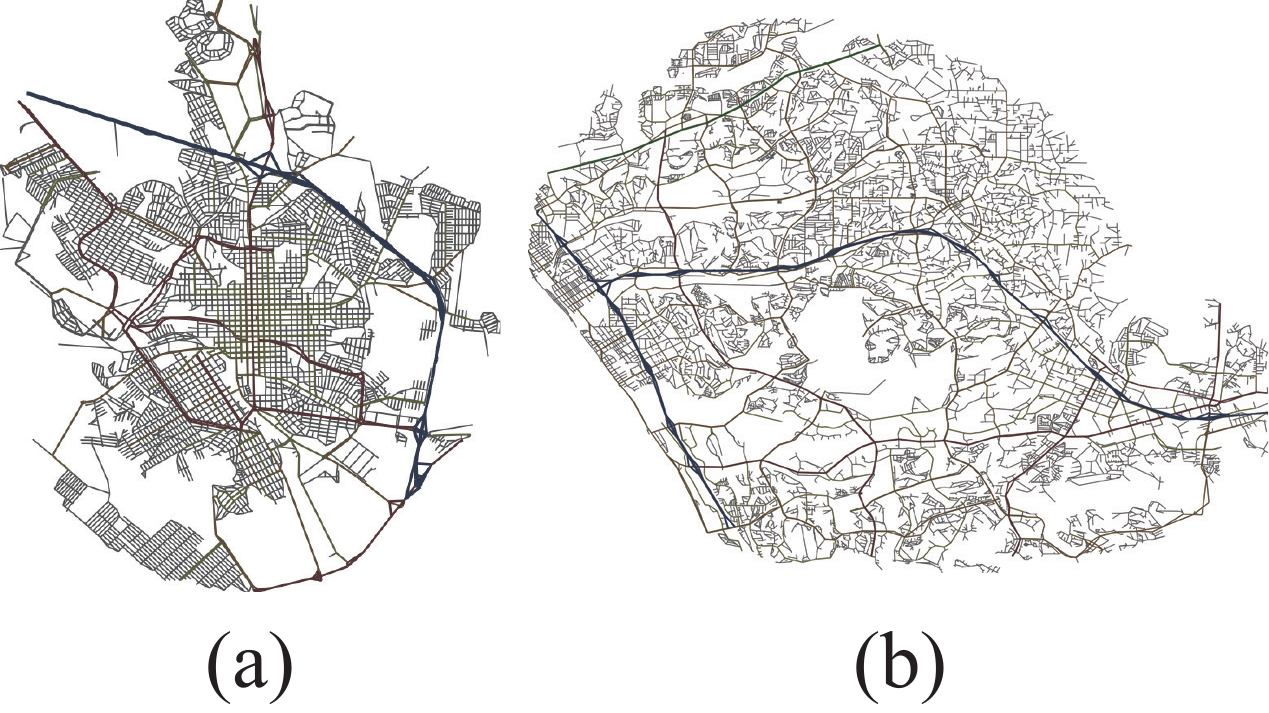}
		\caption{Example of networks obtained by the methodology developed \cite{COMIN2016} considering the urban regions of the cities of Sao Carlos - Brazil (a) and Oceanside - United States (b). The vertices  are arranged according to the original geographical positions of each crossing or termination.}
		\label{RedesUrbanas}
	\end{figure}

The data obtained from the OpenStreetMaps dataset~\cite{OSM2016} can be represented as a network. In this case, each intersection and termination of streets is represented by a vertex, and the presence of a street between two vertices is expressed in terms of an edge. Since the vertices positions are known, we are able to build both the topological and the geometric representation of the streets of a city. Figure~\ref{RedesUrbanas} illustrates two networks of urban regions with different topologies, the former being more regular and lattice-like than the second.

	\subsection{Delimiting the urban regions}

	After obtaining the network representation of the selected cities (see Figure~\ref{f:filter_methodology}), the developed methodology~\cite{COMIN2016} was applied in order to delimit the urban region from each considered city. The methodology consists in defining the \emph{structural density} of the respective city, which is then used for defining the urban region. Specifically, the positions of the network vertices are used as input to a Kernel density Estimation method~\cite{wand1994kernel}, which allows the definition of a probability density function (pdf) for all points in space. This pdf is then associated to the structural density of the city, since larger pdf values are related to more street intersections and terminations. Then, a threshold is applied to the pdf, defining a candidate urban region. Next, the skeleton~\cite{costa2009shape} of the candidate region is obtained, and used to separate urban regions that are close, but have small overlap of streets. The largest remaining region is taken as the urban area of the city.
    
\subsection{Topological properties} \label{propertiesSubsection}
In order to diminish the influence of the network size on its characterization, we adopted only topological properties that are defined locally for the vertices. In particular, the concentric node degree~\cite{ALBERT, COSTA2006}, the concentric clustering coefficient~\cite{ALBERT,COSTA2006}, the accessibility~\cite{travenccolo2008accessibility} and the matching index~\cite{sporns2003graph}\cite{kaiser2004edge}. These measurements are described as follows.
	
	\subsubsection{Concentric measurements}
The concentric measurements~\cite{COSTA2006} extend the traditional measurements of complex networks, such as the node degree and clustering coefficient, so as to account for the structure of the neighborhoods of a vertice. Here, we employed two of these measurements for the analysis: the concentric node degree and concentric clustering coefficient.
	
\begin{figure}[htbp]
		\centering
		\includegraphics[width=\linewidth]{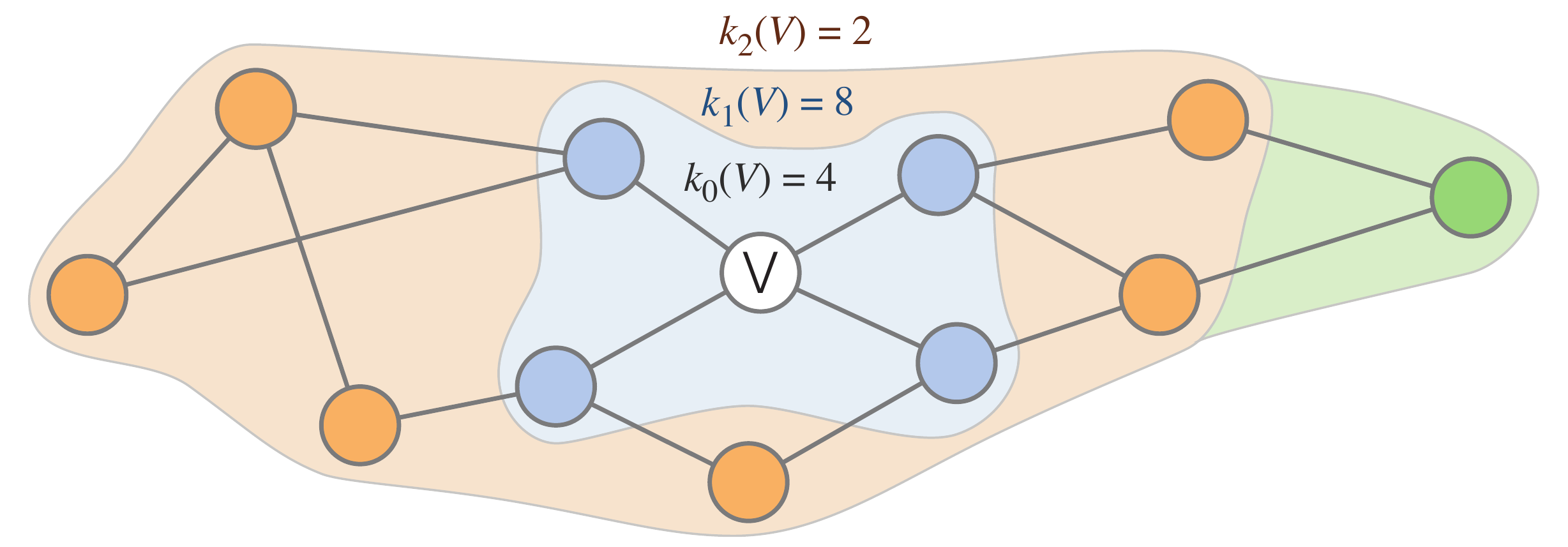}
		\caption{Representation of a graph and the respective neighborhoods of a vertex $V$, corresponding to the concentric levels for $i=1, i=2 and i=3$, where $k_i(V)$ represents the concentric node degree of the neighborhood of level $i$ of vertex $V$.}
		\label{Degree_Neighborhood}
	\end{figure}

	The concentric node degree is a measurement that characterizes the connectivity for successive neighborhoods of a vertex. For a given vertex, the vertices connected to it define a neighborhood of level $1$, the vertices connected to that neighborhood define a neighborhood of level $2$, and so on. The degree of a neighborhood of level $i$ is given by the number of connections between neighborhoods $i$ and $i+1$. Figure~\ref{Degree_Neighborhood} shows an illustration of the concentric degree.
	
The concentric clustering coefficient of a given vertex $V$ is related to the number of connections between the vertices in a given neighborhood. It can be calculated as the ratio of the number, $e_i(V)$, of existing connections between vertices of neighborhood $i$ and the total number of possible connections between these vertices, that is:
	\begin{equation}
		Cc_i(V) = \frac{2 e_i(V)}{n_i(V)(n_i(V)-1)}, 
	\end{equation}
	where $n_i(V)$ is the number of vertices contained in the neighborhood at level $i$.
	For instance, in the example of figure \ref{Degree_Neighborhood} the concentric clustering coefficients are $Cc_1(V) = 0$ and $Cc_2(V) = 0.13$, for, respectively, neighborhoods 1 and 2.

	\subsubsection{Accessibility}
    
The accessibility~\cite{viana2012effective,travenccolo2008accessibility} measures the effective number of vertices that can be accessed from a given vertex $V$ and a given distance $h$. This measurement considers the transition probabilities $P_h(j,V)$ of a random walk starting in vertex $V$ and arriving at target vertex $j$ at distance $h$. The accessibility can be defined as
    \begin{equation}
		A_h(V) = e^{E_h(V)},
	\end{equation}
    where $n$ is the number of vertices having a topological distance $h$ from vertex $V$ and $E_h(V)$ is the entropy signature of vertex $V$ for $h$ steps and is calculated as
    \begin{equation}
		E_h(V) = - \sum_{j=1}^{N}{P_h(j,V) \log{P_h(j,V)}}
	\end{equation}
    
One of the main characteristics of the accessibility is its ability to detect borders in complex networks~\cite{travenccolo2008accessibility}. 
	
	\subsubsection{Matching Index}
The matching index~\cite{sporns2003graph,kaiser2004edge} measures the overlap between the neighborhoods of two connected vertices. This measure can be defined, for a given pair of nodes $i$ and $j$, as the ratio between the number of vertices that are neighbors of both nodes $i$ and $j$ and the total number of connections of $i$ and $j$, that is:
	\begin{equation}
		M(i,j) = \frac{k}{h(i) + h(j) - 2},
	\end{equation}
	where $h(i)$ is the node degree of vertex $i$ and $k$ is the number of vertices connected to both nodes $i$ and $j$. As this is the only measurement referring to the edges and not the vertices, it is interesting to define also the vertex version of this measurement as the average of the matching index of each edge connected to this vertex. This latter type of matching index is adopted henceforth.
	
	\subsection{Framework for city characterization}

\begin{figure}[H]
			\centering
			\includegraphics[width=0.50\linewidth]{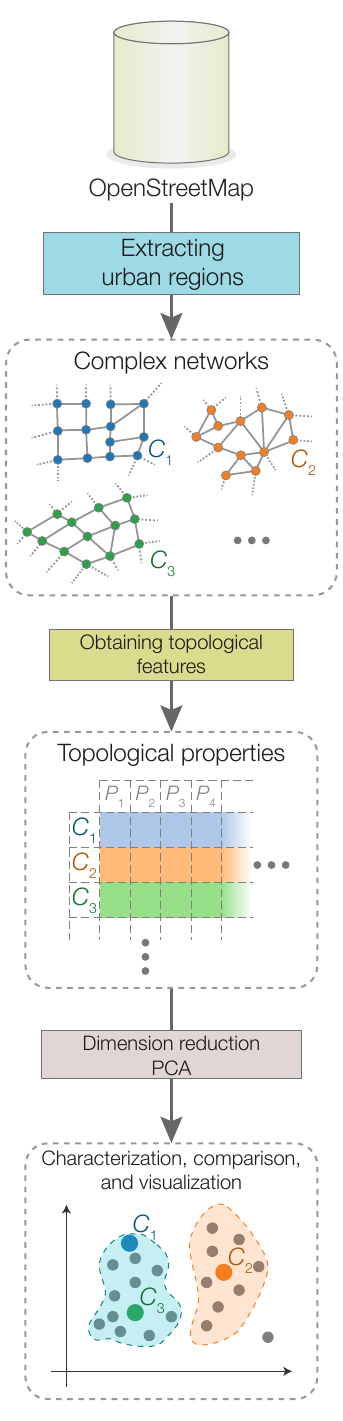}
			\caption{Framework adopted for the analysis of cities.}
			\label{Procedures}
		\end{figure}

    The overall framework used for characterizing the street networks is illustrated in Figure~\ref{Procedures}. 
    
The topological measurements described in Section~\ref{propertiesSubsection} were applied to all street networks considered in this work. For the concentric measurements (node degree, clustering coefficient and accessibility), we considered neighborhoods of level 2 and 5~\cite{ALBERT, COSTA2006}. We then apply the PCA (Principal Component Analysis) method on the obtained measurements to analyze how cities from distinct continents relate one another.

\section{Results and Discussion}

The node measurements described in Section~\ref{propertiesSubsection} were used to characterize the topology of 1150 cities.  In this section we discuss the analysis of such measurements by considering the application of progressively more sophisticate concepts and methods.  More specifically, we start by analyzing the distribution of each measurement, individually. Next, we consider pairwise relationships between such measurements, as quantified by the Pearson correlation coefficient.  Finally, we apply PCA to investigate the overall distribution of cities, seek for eventual clustering, and identify outliers.

\subsection{Measurement-by-Measurement Analysis}
\label{s:mmanalysis}

\begin{figure*}[!htbp]
 \centering
\includegraphics[width=\linewidth]{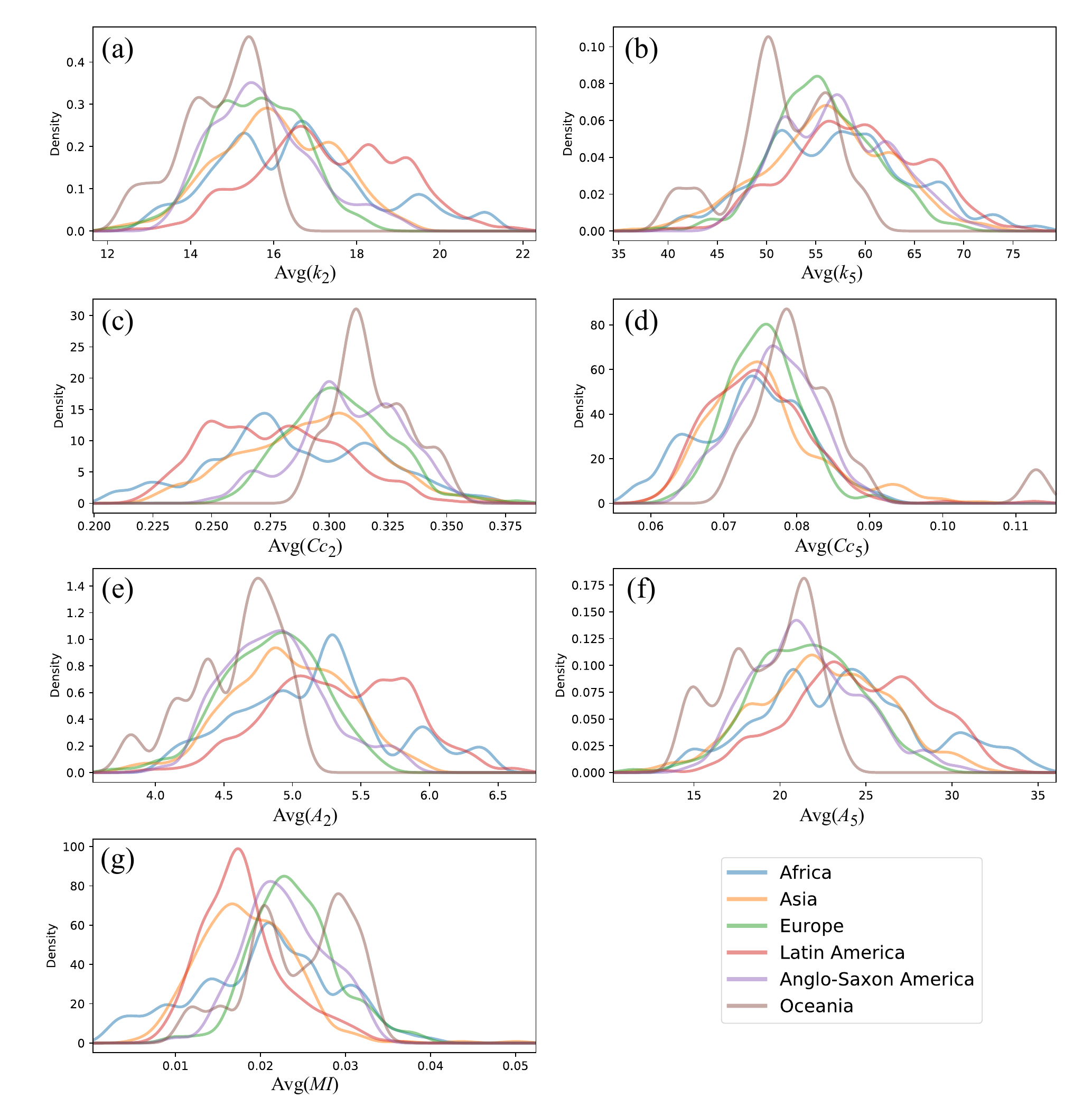}~
\caption{The distributions of the \emph{average} of the considered measurements, with the continents identified by distinct colors.}
		\label{figDensities}
	\end{figure*}

\begin{figure*}[!htbp]
 \centering
 \includegraphics[width=\linewidth]{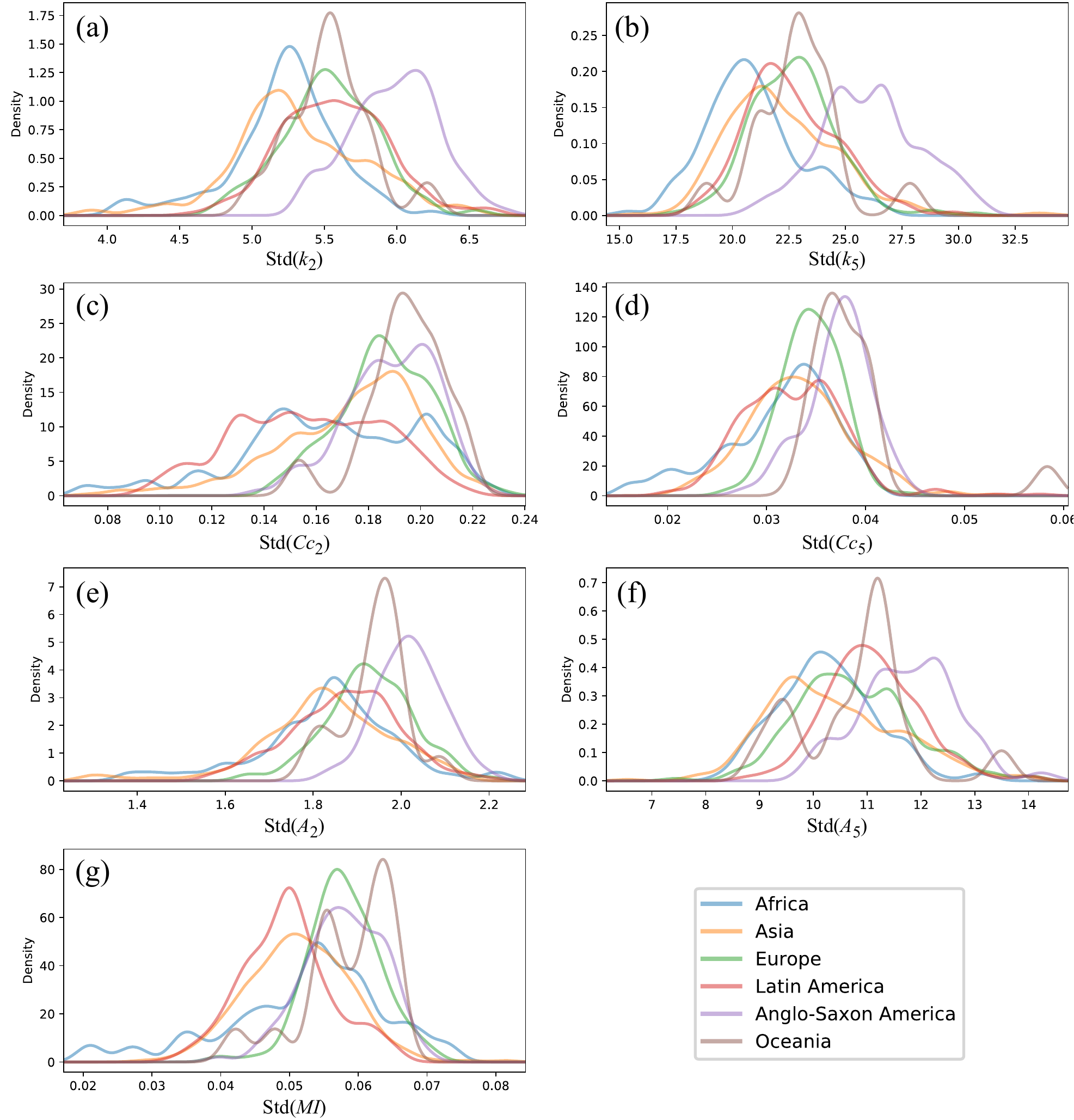}~
\caption{The distributions of the \emph{standard deviation} of the  considered measurements, with the continents identified by distinct colors.}
		\label{figDensities2}
	\end{figure*}

\begin{figure*}[!htbp]
 \centering
\includegraphics[width=\linewidth]{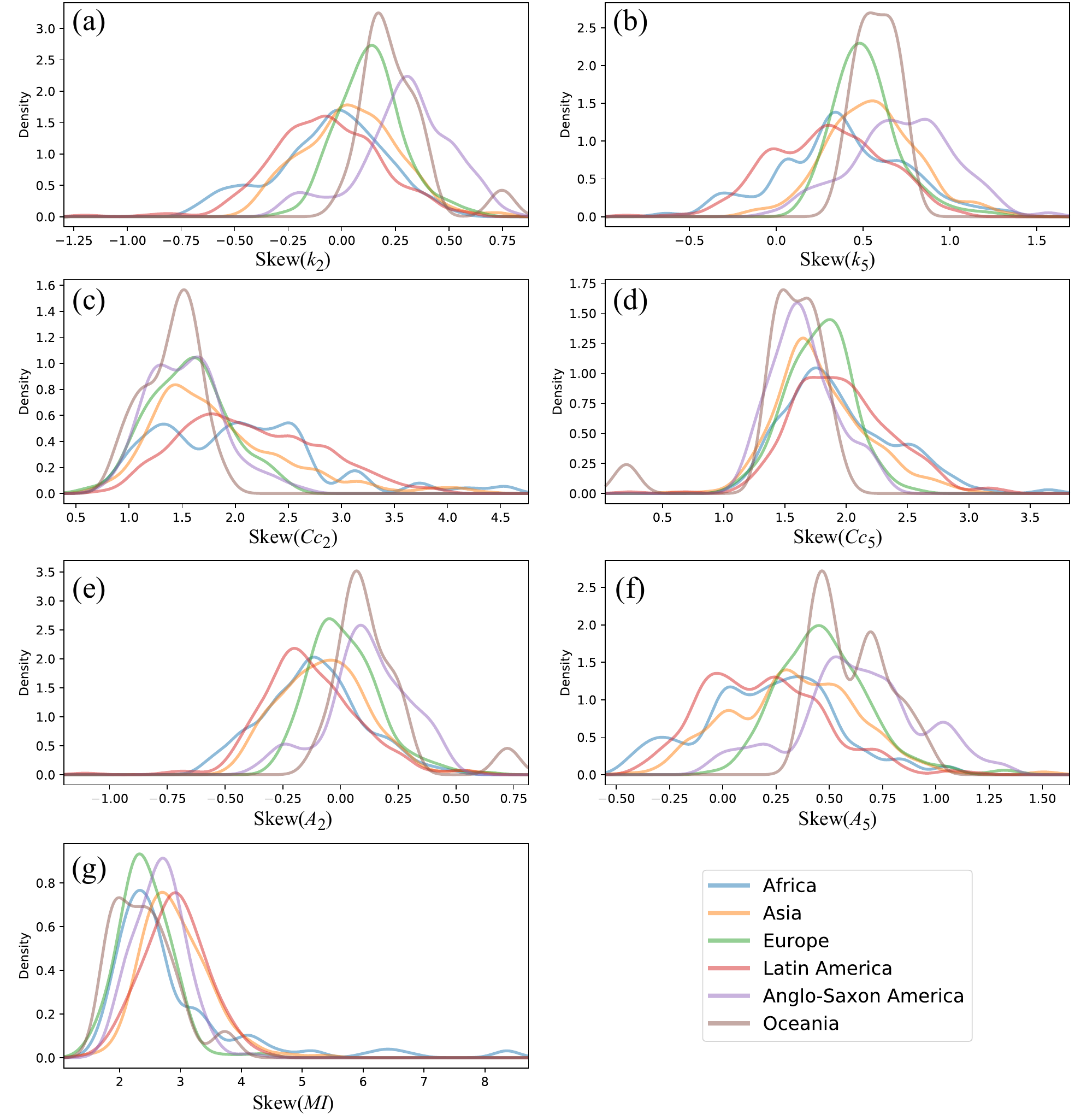}~
\caption{The distributions of the \emph{skewness} of the considered measurements, with the continents identified by distinct colors.}
		\label{figDensities3}
	\end{figure*}

A good starting point in our effort to identify possible differences between urban organization in different continents is to investigate, one-by-one, the distribution of the several considered measurements, which can be divided into three main groups: (a) averages; (b) standard deviations; and (c) skewness for each of the cities.  These measurements are presented in Figures~\ref{figDensities},~\ref{figDensities2} and~\ref{figDensities3} respectively to these three main groups.   Each of these figures is organized with respect to the concentric degree (levels 2 and 5), concentric clustering coefficient (levels 2 and 5), accessibility (levels 2 and 5), and matching index.  

Regarding the concentric degree for level 2 (Fig.~\ref{figDensities}(a)), we have the following order of peaks, ranging from left to right: Oceania, Anglo-Saxo America, Europe, Asia, Africa and Latin-America.  This means that Latin-American cities tend to have, in the average, the larger number of streets crossing one another, being potentially more complex.  Comparable dispersions of this measurement can be observed for the several continents.  Similar trends are observed for the concentric degree of level 5 (Fig.~\ref{figDensities}(b)).

\begin{figure}[!htbp]
		\includegraphics[width=0.4\columnwidth]{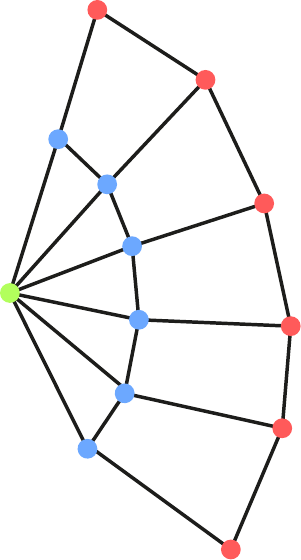}
		\caption{Example of a node, shown in green, associated with a large number of street crossings. The first (blue nodes) and second (red nodes) hierarchies are also shown.}
		\label{f:clusteringIlu}
	\end{figure}

The distributions of concentric clustering coefficient for level 2 are shown in Figure~\ref{figDensities}(c).  The ascending order of peaks observed for this measurement is: Latin-America, Africa, Europe, Anglo-Saxon America, Asia and Oceania.  Interestingly, this ordering is almost the opposite of that verified for the concentric degrees, suggesting that the continents with larger number of street crossings also have less transitivity among such streets. This is illustrated in Figure~\ref{f:clusteringIlu}, which shows a node (colored in green), defined by the crossing of several streets, and the respective blocks belonging to the first two hierarchies of the node.  The impossibility to have connections between non-adjacent corners of such blocks imply in the observed reduction of the clustering coefficient.  The distribution of concentric clustering coefficient for level 5 is shown in Figure~\ref{figDensities}(d).  All peaks shifted to near-zero values, indicating that very little transitivity is observed for higher concentric levels in most cities.  

Figure~\ref{figDensities}(e) shows the distribution of the accessibility for level 2.  These distributions are remarkably similar to the respective concentric degrees for level 2, shown in Fig.~\ref{figDensities}(a), but the accessibility values are generally smaller than the respective degrees.  This indicates that nodes in the second neighborhood tend to be ineffectively accessed, probably as a consequence of block irregularities.  This inefficiency is similar for most Continents.  Similar results are observed with respect to the accessibility for level 5 (Fig.~\ref{figDensities}(f)).

Regarding the standard deviation and skewness results in Figures~\ref{figDensities2} and \ref{figDensities3}, a respective conceptual interpretation is more difficult.  Generally, we have that most of the plots in these figures do not indicate evident displacement of cities from any continent, except for the plots corresponding to Std($k_2$) (Figure~\ref{figDensities2}(a)), Std($k_5$) (Figure~\ref{figDensities2}(b)), Std($A_2$) (Figure~\ref{figDensities2}(e)), Std($A_5$) (Figure~\ref{figDensities2}(f)) and Skew($k_5$) (Figure~\ref{figDensities3}(b)).  These figures indicate a shift of the Anglo-Saxon American cities with respect to the other cities.

\subsection{Correlation Analysis}

\begin{figure}[htbp]
		\includegraphics[width=\linewidth]{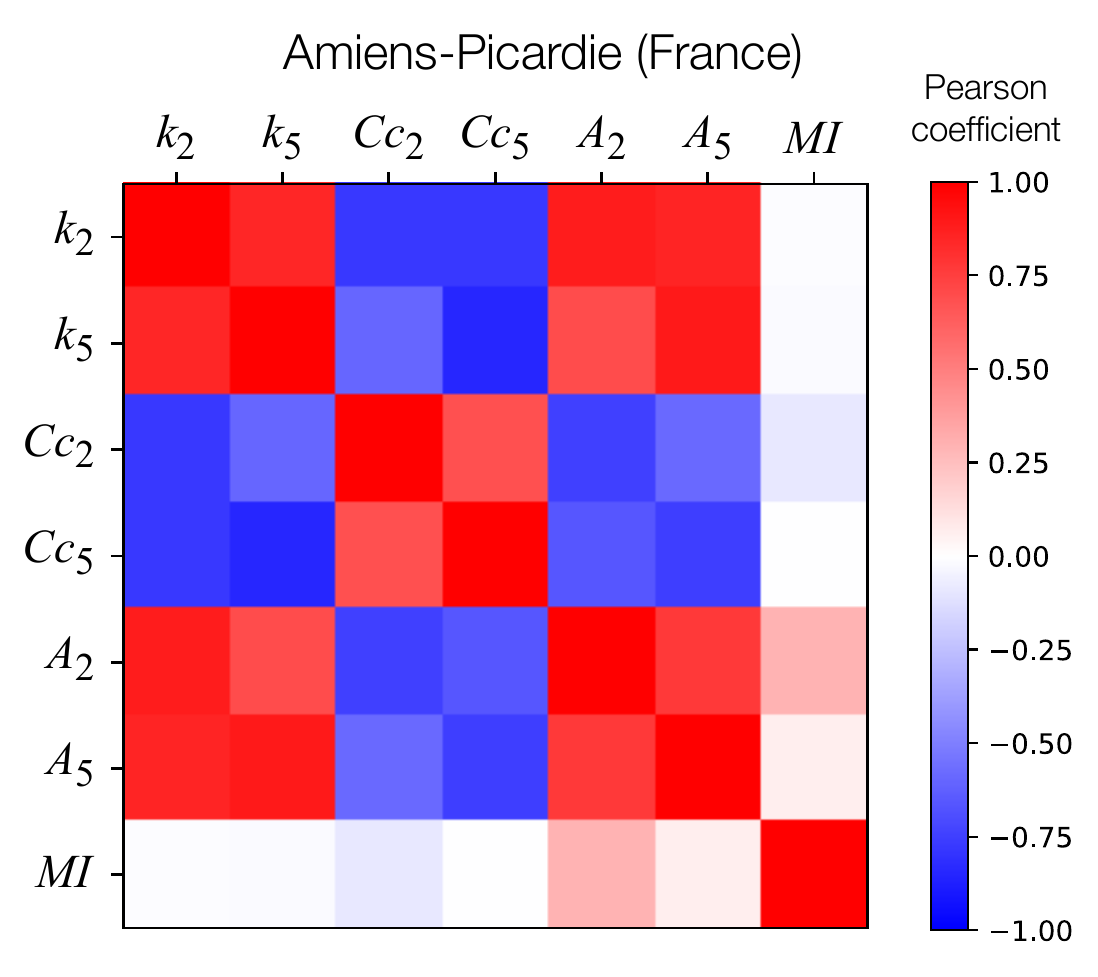}
		\caption{Pearson's correlation matrix for the city of Amiens, France. A strong red color indicates a large Pearson correlation while a strong blue color is associated with negative Pearson correlation. The labels on the axes correspond to: $k_i$ - concentric node degree of neighborhood of level $i$; $Cc_i$ - concentric clustering coefficient of neighborhood of level $i$; $A_i$ - accessibility of neighborhood of level $i$; $MI$ - matching index.}
		\label{correlacao1}
	\end{figure}

The relationship between the adopted measurements applied to each node can be quantified by considering the Pearson correlation coefficient between all possible pairwise combinations of them. The respective results can be organized as a matrix, henceforth called \emph{correlation matrix}. We estimated these matrices for each of the considered cities.  A typical correlation matrix is shown in Figure~\ref{correlacao1}. In this figure, we see that corresponding measurements calculated at different scales are highly correlated among themselves. Also, the concentric degree tends to be positively correlated with accessibility. This means that, generally speaking, lower degrees are usually related to larger irregularities in the topology of city networks. The concentric clustering coefficient is negatively correlated with the concentric degree and accessibility. The matching index is uncorrelated with the previously mentioned measurements. 

\subsection{PCA}

\begin{figure*}[!htbp]
		\includegraphics[width=\linewidth]{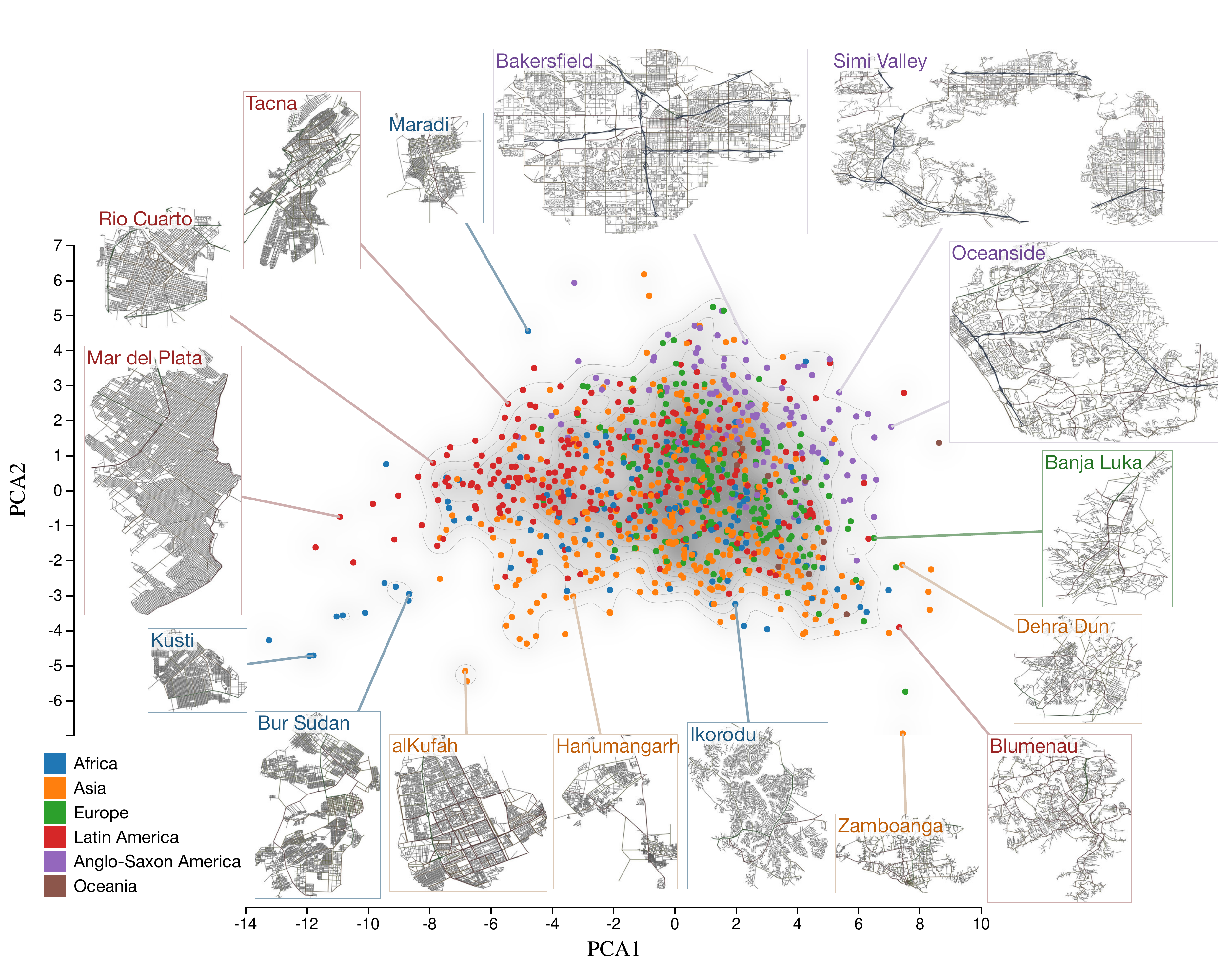}
		\caption{PCA generated from the assembled database with maps of some cities highlighted. The explained variation of the first two components PCA1 and PCA2 are $0.59$ and $0.17$, respectively.}
		\label{PCA_cidades}
	\end{figure*}

For each city, we calculated the average, standard deviation and skewness~\cite{altman1996,oja1983} of the considered measurements, thus defining a total of 21 measurements to characterize the cities. We then applied PCA (Principal Component Analysis) to project the data into two dimensions, obtaining the result shown in Figure~\ref{PCA_cidades}. In this figure, each point represents a city, being colored according to the continent where the city is located. A number of interesting trends can be observed in this plot: i) a cluster of cities from Latin America is observed on the left-hand side of the PCA space; ii) cities from Anglo-Saxon America tend to fall on the upper right-hand side; iii) cities from Europe and Asia tend to occupy the lower right-hand side of the PCA space, also presenting a large overlap, but the European cities resulted more tightly clustered than cities in Asia. 

Each PCA axis is defined as a linear combination of the original measurements. Therefore, we can look at the coefficients of the linear combination to identify the measurements that had the largest influence on the PCA projection. Such weights are shown in Table \ref{Table1}.

\begin{table*}[]
\caption{Weights of influence on the PCA projection for each measure.}
\centering

\label{Table1}

\resizebox{\textwidth}{!}{
\begin{tabular}{c|c c c c c c c c c c c }
  Comp. & Avg($k_2$) & Std($k_2$) & Skew($k_2$) & Avg($k_5$) & Std($k_5$) & Skew($k_5$) & Avg($Cc_5$) & Std($Cc_5$) & Skew($Cc_5$) & Avg($Cc_2$) & Std($Cc_2$) \\
  \hline
PCA1 & 0.27 & -0.10 & -0.26 & 0.24 & -0.07 & -0.20 & -0.24 & -0.25 & 0.26 & -0.27 & -0.26 \\
PCA2 & -0.14 & -0.46 & -0.05 & -0.23 & -0.45 & 0.02 & 0.13 & -0.08 & -0.04 & 0.05 & -0.01
\end{tabular}}

\vspace*{0.25cm}

\resizebox{\textwidth}{!}{
\begin{tabular}{c|c c c c c c c c c c }
  Comp. & Skew($Cc_2$) & Avg($A_2$) & Std($A_2$) & Skew($A_2$) & Avg(acces5) & Std($A_5$) & Skew($A_5$) & Avg($MI$) & Std($MI$) & Skew($MI$)\\
  \hline
PCA1 & 0.26 & 0.27 & -0.16 & -0.25 & 0.27 & -0.00 & -0.26 & -0.12 & -0.15 & 0.12\\
PCA2 & 0.02 & -0.13 & -0.38 & -0.04 & -0.17 & -0.50 & -0.02 & -0.09 & -0.09 & 0.13\\

\end{tabular}}
\end{table*}

Regarding PCA1, i.e. the first principal axis, similar weights were obtained for the greatest part of the measurements, suggesting that they have similar roles in the distribution of the cities.  A different situation was verified with respect to PCA2, with four measurements (the standard deviations of the accessibility for levels 2 and 5, and the standard deviations of concentric degree for concentric levels 2 and 5) having larger weights. However, such measurements are difficult to be gauged by visual inspection of the cities topologies.  Given such issues, it is difficult to identify, intuitively, 
which topological features of the cities vary along the PCA1 and PCA2 axes.

\begin{figure*}[!htbp]
		\includegraphics[width=\linewidth]{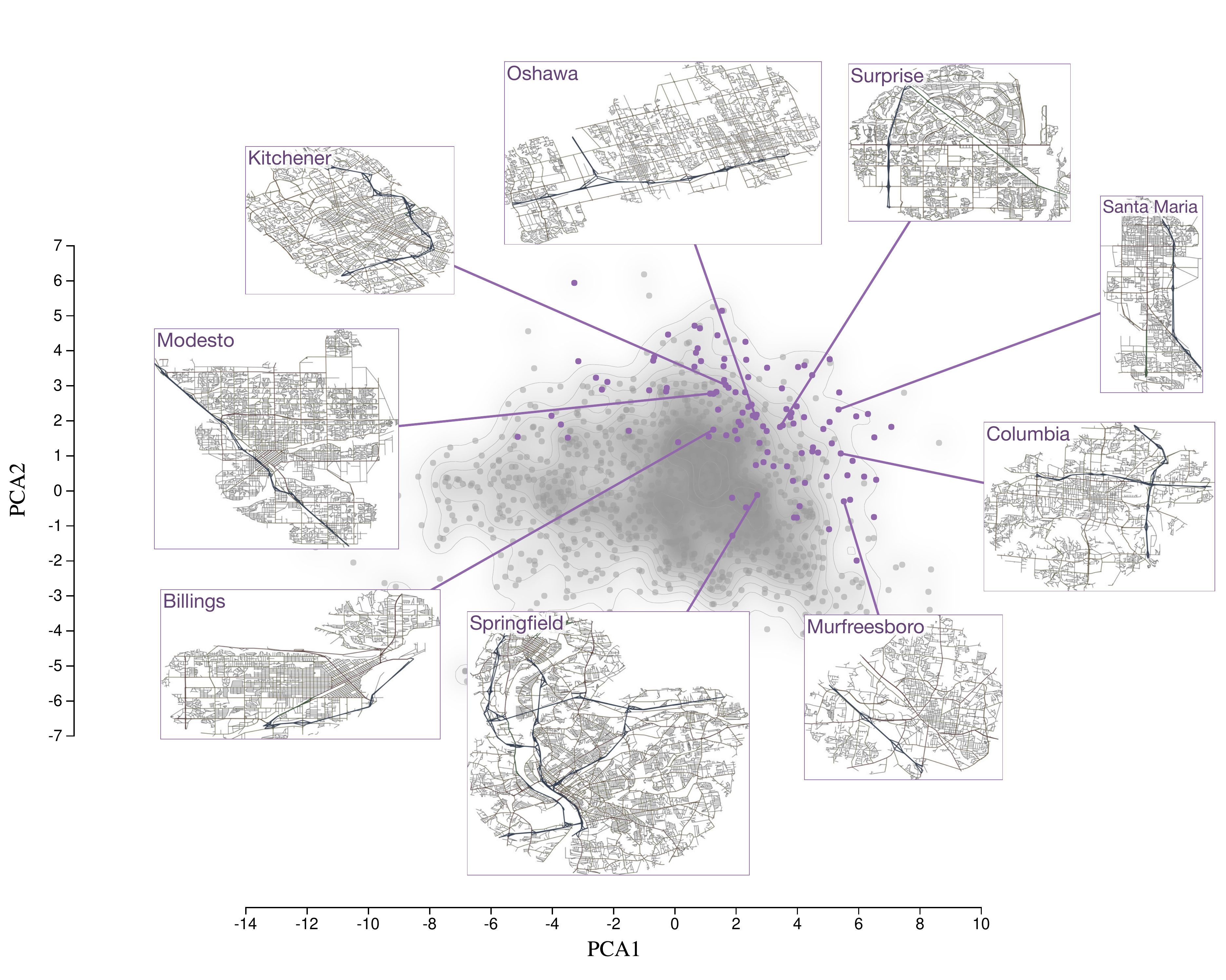}
		\caption{PCA generated from the assembled database. Points from Anglo-Saxon America are shown in purple.}
		\label{angloSaxonAmerica}
	\end{figure*}

Figure~\ref{angloSaxonAmerica} shows the cities obtained for Anglo Saxon America, which yielded a respective PCA distribution with the smallest overlap with regions defined by the other continents.  Recall that the cluster obtained for Anglo Saxon America occupies the upper right-hand side of the overall PCA. As can be inferred from analysis of Fig.~\ref{angloSaxonAmerica}, the cities in this continent seem to present, with a few exceptions, a more regular global pattern of connections, organized along mostly orthogonal or oblique axes.  Additional insight about the nature of the shift presented by the cities in Anglo-Saxon American can be gained by referring to the measurement-by-measurement results presented and discussed in Section~\ref{s:mmanalysis}.   Remarkably, the distributions of the vast majority of the measurements obtained for this continent resulted in the middle of the other continents, accounting for no shifting of the Anglo-Saxon cities.  This shift is only related to 5 out of the 21 measurements in Figures~\ref{figDensities},~\ref{figDensities2} and~\ref{figDensities3}, which show the Anglo-Saxon cities displaced from the other cities.  Interestingly, the PCA led to a visualization that emphasized this displacement.

	\section{Conclusions}
    The analysis of street structure in cities has great importance for the understanding of aspects such as population density and traffic dynamics and can be used to improve the traffic system itself or provide subsidies for city planing. This analysis is also useful to differentiate cities, according to their topological properties, in order to classify them and to understand the potential effects from their historical or geographical formation.
	
	In this work, we considered 1150 world cities, which had their region or interest identified and were then represented as complex networks. Several topological measurements were calculated and discussed one-by-one, according to pairwise relationships, and studied as continent-related groups in PCA projections.  Through these analyzes, it was possible to identify patterns in the networks, such as the tendency of nodes with higher degree to have smaller clustering coefficient. It was also possible to identify continent specific topological trends, especially for Anglo-Saxon America. Interestingly, this separation could be observed in just 5 out of the 21 considered measurements.  Nevertheless, the visualization obtained by PCA was able to emphasize the displacement of the Anglo-Saxon cities.
 
The reported approach and results pave the way to several future studies, such as the integration with other networks (e.g. train, roads, underground), incorporation of dynamics for modeling human displacement, and consideration of additional properties of nodes (e.g.~ presence of traffic lights, zebra crossings, etc.) and edges (arc length, number of lanes, maximum speed, etc.).

\acknowledgements    
G. S. Domingues thanks CNPQ (Grant No. 2016-606) for financial support. F. N. Silva acknowledges FAPESP (Grant No. 15/08003-4). C. H. Comin thanks FAPESP (Grant No. 15/18942-8) for financial support. L. da F. Costa thanks CNPq (Grant No. 307333/2013-2) for support. This work has also been supported by the FAPESP grant 11/50761-2.  The authors also thank Ema Strano for interesting suggestions.
    
	\bibliographystyle{apa}
	\bibliography{referencias}
\end{document}